\documentclass[11pt]{article}
\pdfoutput=1
\usepackage{jheppub,amsmath,amssymb}
\usepackage{booktabs}% http://ctan.org/pkg/booktabs
\usepackage{hyperref}
\usepackage{graphicx}
\usepackage{mathtools}
\usepackage{xcolor}
\usepackage{color}
\usepackage{amsfonts}
\usepackage{mathtools}
\usepackage{amssymb}
\usepackage{array}
\usepackage{comment}

\usepackage{dcolumn}% Align table columns on the decimal point
\usepackage{bm}% bold math
\usepackage{braket}
\usepackage{microtype} %makes everything nicer

\numberwithin{equation}{section}

\newcommand{\ud}{\mathrm{d}}

\newcommand{\Dop}{\mathcal{D}}

\newcommand{\mn}{{\mu\nu}}
\newcommand{\mnrs}{{\mu \nu \rho \sigma}}

\newcommand{\blue}[1]{ \textcolor{blue}{#1}}

\newcommand{\hone}{h_{vv}}
\newcommand{\htwo}{h_{v\bar{w}}}

\def\be{\begin{equation}}
\def\ee{\end{equation}}
\newcommand{\eqs}[1]{\begin{align}#1\end{align}}

\begin{document}
\title{Self-Dual Cosmology}
	\author[a]{Mariana Carrillo Gonz\'alez,}  
	\affiliation[a]{Theoretical Physics, Blackett Laboratory, Imperial College, London, SW7 2AZ, U.K }
    \author[b]{Arthur Lipstein,}
    \author[b]{Silvia Nagy}
    \affiliation[b]{Department of Mathematical Sciences, Durham University, Durham, DH1 3LE, UK}

    \emailAdd{m.carrillo-gonzalez@imperial.ac.uk}
    \emailAdd{arthur.lipstein@durham.ac.uk}
    \emailAdd{silvia.nagy@durham.ac.uk}

\abstract{
We construct cosmological spacetimes with a self-dual Weyl tensor whose dynamics are described by conformally coupled scalars with only cubic self-interactions. Similar to the previously discovered cases in flat and (Anti) de Sitter backgrounds, the interactions are characterized by a bracket that encodes a kinematic algebra. We discuss how the color-kinematics duality and double copy are realized in these cosmological backgrounds. If we further impose that the Ricci scalar is that of an FLRW spacetime, we find two new self-dual metrics corresponding to radiation-dominated and coasting (non-accelerating) FLRW backgrounds. Relaxing this requirement, we find an infinite family of solutions given by three different conformal classes of cosmological self-dual metrics. These solutions approximate those of FLRW as long as we impose a simple additional constraint on the scalar theory.  
}

\maketitle

\pagebreak

\section{Introduction}
Computations in curved spacetimes can be extremely involved, mainly due to the high non-linearity of Einstein's equations. However, this same non-linearity allows for exciting solutions to arise in gravitational theories. When working on asymptotically flat spacetimes, simplifications can arise. For example, the scattering amplitudes of gravitons can be written as the double copy of the scattering amplitudes of gluons \cite{Bern:2008qj, Bern:2010ue, Bern:2010yg, Kawai:1985xq}. This can be done by exchanging color structures with kinematic ones, which satisfy the same algebra due to the color-kinematics duality. One of the simplest realizations of the double copy is within the so-called self-dual sector \cite{Plebanski:1975wn,Bardeen:1995gk,Chalmers:1996rq,Prasad:1979zc,Dolan:1983bp,Parkes:1992rz,Cangemi:1996pf,Popov:1996uu,Popov:1998pc,Skvortsov:2024rng}. This corresponds to the sub-sector of solutions that have a self-dual curvature tensor.

It is well known that Ricci flat spacetimes that are solutions to the vacuum Einstein equations with a self-dual Riemann tensor can be described by a single scalar satisfying the Plebanski ``heavenly'' equation \cite{Plebanski:1975wn}. Furthermore, when working in the lightcone gauge, the interactions of this scalar can be written in terms of nested Poisson brackets acting in a two-dimensional subspace, which gives rise to a single cubic vertex. This cubic vertex corresponds to the $(++-)$ vertex. For real momenta, all the tree-level scattering amplitudes of this theory vanish, which is a consequence of their classical integrability \cite{Ward:1977ta,Dunajski_1998,Prasad:1979zc,Dolan:1983bp,Popov:1996uu,Popov:1998pc,Park:1989vq,Husain:1993dp,Bardeen:1995gk}. This property is broken at loop order, but the amplitudes are simple rational functions \cite{Bern:1998sv}, and the theory is one-loop exact, since no 2-loop or higher diagrams can be written with the single  $(++-)$ vertex. Similar statements hold for self-dual Yang-Mills in a flat spacetime. In this case, the interactions are given in terms of a Lie bracket and the same Poisson bracket as in the gravitational case. This structure showcases one of the simplest examples of color-kinematics duality. The Poisson bracket encodes the kinematic algebra, which in this case is given by area-preserving diffeomorphisms \cite{Monteiro:2011pc}. Different realizations of the double copy in the self-dual sector have been described in \cite{Chacon:2020fmr, Campiglia:2021srh, Monteiro:2022nqt, Borsten:2023paw, Armstrong-Williams:2022apo, Skvortsov:2022unu, Elor:2020nqe, Farnsworth:2021wvs,Berman:2018hwd,Nagy:2022xxs,Bonezzi:2023pox,Diazinprep,He:2015wgf,Easson:2023dbk,Brown:2023zxm,Doran:2023cmj}. Other explicit realizations of kinematic algebras or Lagrangians with explicit color-kinematics duality have arisen in various contexts for theories beyond the Yang-Mills self-dual sector \cite{Monteiro:2013rya,Cheung:2016prv,Chen:2019ywi,Borsten:2023ned,Ben-Shahar:2021zww,Ben-Shahar:2022ixa,Brandhuber:2021bsf,Brandhuber:2022enp,Fu:2016plh,Reiterer:2019dys,Tolotti:2013caa,Ferrero:2020vww,Borsten:2019prq,Borsten:2020zgj,Borsten:2020xbt,Borsten:2021hua,Diaz-Jaramillo:2021wtl,Bonezzi:2022bse,Bonezzi:2022yuh,Bonezzi:2023lkx,Bonezzi:2024dlv,Armstrong-Williams:2024icu}.

An interesting question that has been asked in the past few years is whether a double copy relation can exist in curved spacetimes, and hence help us to simplify the complicated calculations arising in these backgrounds. This has been explored in many different contexts in \cite{Carrillo-Gonzalez:2017iyj,Bahjat-Abbas:2017htu,Farrow:2018yni,Lipstein:2019mpu,Borsten:2019prq,Albayrak:2020fyp,Armstrong:2020woi,Prabhu:2020avf,Sivaramakrishnan:2021srm,Alkac:2021bav,Jain:2021qcl,Zhou:2021gnu,Alday:2021odx,Borsten:2021zir,Armstrong:2022csc,Diwakar:2021juk,Han:2022mze,Cheung:2022pdk,Herderschee:2022ntr,Drummond:2022dxd,Lipstein:2023pih,Armstrong:2023phb,Farnsworth:2023mff,Mei:2023jkb,Liang:2023zxo,Ilderton:2024oly,Adamo:2024hme}, but a systematic understanding of the double copy in general backgrounds is still lacking. A promising avenue is to consider self-dual theories in curved spacetimes. A first step in this direction was taken in \cite{Lipstein:2023pih}, where self-dual gravity in Anti-de Sitter (AdS) space was reduced to a simple cubic scalar theory which arises from the double copy of self-dual Yang-Mills in AdS, and exhibits a deformed $w_{1+\infty}$ algebra analogous to that of self-dual gravity in flat spacetime \cite{Fairlie:1990wv,Strominger:2021mtt}\footnote{See also \cite{Pope:1991ig,Monteiro:2022lwm,Pope:1989ew,Bu:2022iak,Bittleston:2023bzp,Bittleston:2024rqe,Taylor:2023ajd,Kmec:2024nmu,Himwich:2021dau,Adamo:2021lrv,Nagy:2024dme,Nagyinprep} for further work and deformations of these algebras in gravity and YM.}. Other formulations of self-dual gravity in de Sitter (dS) space were constructed in \cite{Krasnov:2016emc,Krasnov:2021cva,Neiman:2023bkq,Neiman:2024vit}. The $w_{1+\infty}$ symmetry in AdS was subsequently studied from other perspectives in \cite{Taylor:2023ajd,Bittleston:2024rqe}. In this paper, we will explore a further generalization of these ideas to cosmological spacetimes. In particular, we show that they extend to self-dual gravity in radiation dominated and coasting (non-accelerating) FLRW spacetimes, as well as an infinite class of solutions, obtained by performing Weyl transformations, whose stress tensors become FLRW-like after imposing a certain constraint on the scalar theory. The scalar theory describing self-dual gravity in these backgrounds contains cubic interactions constructed from Jacobi brackets \cite{Lichnerowicz1977LesVD,JacobiVaisman,cabauBook}, which encode a kinematic algebra analogous to the color algebra of self-dual Yang-Mills, reflecting a color/kinematics duality in these backgrounds.    

The outline of the paper is as follows. In Section \ref{sec:SDgravity}, we introduce the concept of off-shell and on-shell self-dual Weyl tensor. We show that the on-shell case, where the Ricci tensor is fully fixed by a chosen stress-energy tensor, only gives rise to the known flat and (A)dS cases. These self-dual solutions can be cast in terms of a scalar theory with cubic interactions. We extend this description to other backgrounds by considering solutions to the self-dual Weyl tensor without incorporating the Einstein equations, that is, in the off-shell case. This construction gives three conformal classes of metrics with self-dual off-shell Weyl tensors. For each of these conformal classes, we find a Jacobi bracket that characterizes their cubic interactions. We continue in Section \ref{sec:DoubleCopy} by deriving these self-dual solutions from the double copy of Yang-Mills in conformally flat backgrounds and showing that they exhibit a deformed $w_{1+\infty}$ algebra which is closely tied to the kinematic algebra encoded by the Jacobi brackets. This construction generalises the previously known flat and dS cases to more general cosmological self-dual solutions. In Section \ref{sec:sdCosmology}, we analyse the stress tensors and equations of state of the new cosmological self-dual solutions to gain further insight into their physical interpretation.
We highlight two new interesting cases whose Ricci scalar is that of an FLRW metric; these are radiation-dominated and coasting FLRW self-dual solutions.
Finally, we summarize our results and discuss future directions in Sec.~\ref{sec:Concl}. In the Appendices, we briefly review power-law cosmologies and provide more details about the self-dual Weyl tensor equation, Jacobi brackets, and properties of the new self-dual solutions.

%%%%%%%%%%%%%%%%%%%%
\section{Self-dual gravity in the presence of sources} \label{sec:SDgravity}

In spacetimes with vanishing Ricci tensor, the condition of a self-dual Weyl tensor is reduced to 
\be \label{sd_cond_flat}
R_{\mu \nu \rho\sigma} = \tfrac{1}{2} \epsilon_{\mu \nu}^{\phantom{\mu \nu }\eta \lambda} R_{\eta \lambda\rho\sigma}.
\ee 
where $\epsilon_\mnrs=\sqrt{g}\varepsilon_\mnrs$, $\varepsilon_\mnrs$ the 4-dimensional Levi-Civita symbol, and we work in Euclidean signature\footnote{We can obtain non-trivial solutions in Lorentzian signature by adding a factor of $i$ in the right-hand side of Eq.~\eqref{sd_cond_flat} and considering complex solutions.}. Note that this equation encodes both the Einstein equations and the Bianchi identity when contracting two of its indices. Thus, solving Eq.~\eqref{sd_cond_flat} is enough to find a self-dual solution to the vacuum Einstein equations. 

Let us instead consider the following constraint:
\begin{equation}
  C_\mnrs=\frac{1}{2} {\epsilon_{\mu \nu}}^{\eta \lambda} C_{\eta \lambda \rho \sigma},   \label{eq:SDgravity}
\end{equation}
where the Weyl tensor is given by
\begin{equation}
   {C_{\mu \nu}}^{\rho \sigma}={R_{\mu \nu}}^{\rho \sigma}-2 {R_{[\mu}}^{[\rho} {g_{\nu]}}^{\sigma]}+\frac{1}{3} R {g_{[\mu}}^{[\rho} {g_{\nu]}}^{\sigma]} \ . \label{eq:Weyl}
\end{equation}
Note that this equation is by definition invariant under Weyl transformations. Hence, if we find one solution then we can obtain an infinite family of solutions by applying Weyl transformations. We will refer to this as a conformal class. Recall that the Ricci tensor and scalar are determined by the Einstein equation as follows
\begin{equation}
    R_{\mu\nu}=T_{\mu\nu} - \frac{1}{2}T g_{\mu\nu} \ , \label{eq:EE}
\end{equation}
where $T^{\mn}$ is the stress-energy tensor, and $T\equiv {T_\mu}^\mu$ is its trace. Making this replacement in \eqref{eq:Weyl} then gives an object that we will refer to as the on-shell Weyl tensor: 
\begin{equation}
    \left.{C_{\mu\nu}}^{\rho\sigma}\right|_{{\rm on-shell}}\equiv{R_{\mu\nu}}^{\rho\sigma}-2{T_{[\mu}}^{[\rho}{g_{\nu]}}^{\sigma]}+\frac{2}{3}T{g_{[\mu}}^{[\rho}{g_{\nu]}}^{\sigma]} \ . \label{eq:SDtensor}
\end{equation}
We may then impose self-duality of the on-shell Weyl tensor, which we will refer to as on-shell self-duality:
\begin{equation}
\left.C_{\mnrs}\right|_{{\rm on-shell}}=\frac{1}{2}{\epsilon_{\mu\nu}}^{\eta\lambda}\left.C_{\eta\lambda\rho\sigma}\right|_{{\rm on-shell}}
\label{onshellsd}
\end{equation}
Contracting this equation on both sides with $g^{\nu \sigma}$ then implies the Einstein equations sourced by a generic stress-energy tensor:
\begin{equation}
    R_{\mu \rho}-T_{\mu\rho}+\frac{1}{2} T g_{\mu \rho}=\frac{1}{2} \sqrt{-g} \epsilon_\mu^{\sigma \eta \lambda} R_{\eta \lambda \rho \sigma}=0 \ .
\end{equation}
As in the vacuum case, the left-hand side gives the trace reversed Einstein equations, and the right-hand side gives the Bianchi identity. If we set $T^{\mn}=-\Lambda g^\mn$, we will recover the result for an (A)dS spacetime found in \cite{Lipstein:2023pih}. Note that the on-shell self-duality constraint in \eqref{onshellsd} is not Weyl invariant. 

\subsection{On-shell self-duality} \label{sec:SDflat}
First, we will consider solutions to the on-shell self-dual equations in Eq.~\eqref{onshellsd}. We will work in double lightcone coordinates 
\eqs{
u&=t+iz \ , \quad v=t-iz \ ,\\
w&=x+iy \ , \quad \bar{w}=x-iy \ ,
\label{lccoords}
}
and consider the metric ansatz given by
\be \label{eq:SDfrw}
ds^2=a(\tau)^2 \left(dw\,d\bar{w}-du\,dv + h_{\mu\nu}\,dx^\mu dx^\nu\right) \ ,
\ee
where $\tau$ is the conformal time given by
\be
\tau=(u+v)/2 \ .
\ee
This metric reduces to an FLRW metric when $h_{\mu\nu}=0$. We will refer to this as the background metric. For convenience, we will split the spacetime coordinates as $x^i=(u,w),\ y^\alpha=(v,\bar{w})$. Since our coordinates are complex, $x^i$ corresponds to the holomorphic and $y^\alpha$ the anti-holomorphic sector. We work in lightcone gauge,  $h_{u\mu}=0$, and take the ansatz.
\be
h_{i\mu}=0,\quad h_{\alpha \beta} = \frac{1}{4}\Dop_{(\alpha }\tilde{\Dop}_{\beta)}\phi \label{eq:hphi} \ , 
\ee
where $\Dop_\alpha$ and $\tilde{\Dop}_\alpha$ are differential operators that will be unspecified for now. 

We proceed by making the following assumptions:
\begin{enumerate}
    \item The $\Dop_\alpha$ and $\tilde{\Dop}_\alpha$ operators are at most first order in derivatives and can only depend on functions of conformal time.
    \item When $a\blue{(\tau)}\rightarrow1$, the self-dual solution reduces to the standard result in a flat background were
    \be
    \Dop_\alpha=\tilde{\Dop}_\alpha=(\partial_w,\partial_u) \ ,
    \ee
    and the scalar equation of motion becomes
    \be
     \Box_{\mathbb{R}^{4}} \phi - \{\{\phi,\phi\}\}=0 \,
     \label{scalareomflat}
    \ee
    where the Poisson brackets are defined as 
    \be
    \{f,g\}=\partial_{w}f\partial_{u}g-\partial_{u}f\partial_{w}g\  \ .\label{poisson_def}
    \ee
    The interactions of this scalar are purely on the holomorphic sector and encode the kinematic algebra corresponding to area-preserving diffeomorphisms of the holomorphic $x^i$ plane \cite{Monteiro:2011pc}.
    \item The equation of motion for the scalar contains at most second-order derivatives.
\end{enumerate}

Given these assumptions, we will show that the on-shell self-duality in \eqref{onshellsd} can only be solved in flat or (A)dS backgrounds. We start by looking at the two components of Eq.~\eqref{onshellsd} that involve only $\hone$ and $\htwo$:
\eqs{
 &\left.C_{vuvu}\right|_{{\rm on-shell}}-\frac{1}{2}{\epsilon_{vu}}^{\eta\lambda}\left.C_{\eta\lambda vu}\right|_{{\rm on-shell}}= \nonumber \\
 &\frac{1}{2}a'^2 \hone+a a' \partial_u\hone-a^2\left(\partial_u\partial_w\htwo-\partial_u^2\hone\right) =0 \ ,  \label{eq:onshell1} \\
&\left.C_{\bar{w}vwu}\right|_{{\rm on-shell}}-\frac{1}{2}{\epsilon_{\bar{w}v}}^{\eta\lambda}\left.C_{\eta\lambda wu}\right|_{{\rm on-shell}}=\nonumber \\
&\quad \quad \quad \quad \frac{1}{2}a'\left(a' \hone +a\left(\partial_u\hone-\partial_w\htwo\right)\right) =0 \ .\label{eq:onshell2}
}
Here, prime denotes a conformal time derivative:  $'\equiv\partial_\tau$. The stress tensor in Eq.~\eqref{eq:SDtensor} corresponds to a perfect fluid sourcing an FRW spacetime with scale factor $a$: 
\be
T^\mn=\rho u^\mu u^\nu+ P \gamma^\mn \ , \quad \rho=3\left(\frac{a'}{a^2}\right)^2 \, \quad P=\frac{\rho}{3}-2\frac{a''}{a^3} \ , \label{eq:TmunuPerfectFluid}
\ee
where $\rho$ is the energy density, $P$ is the pressure, $u^\mu$ is a timelike unit vector giving the direction of flow of the fluid, and $\gamma_{\mu\nu}=g_{\mu\nu}+u_\mu u_\nu$ is the metric of the surface perpendicular to the flow. Under assumptions 1-2, we find that to solve Eq.~\eqref{eq:onshell2} we need to fix the operators as
\be
\Dop_\alpha=(\partial_w,\partial_u+F(\tau)) \quad \tilde{\Dop}_\alpha=(\partial_w,\partial_u+\frac{2a'}{a}-F(\tau)) 
\ee
with an arbitrary function $F$. Using these operators on Eq.~\eqref{eq:onshell1} then gives
\be
\left.C_{vuvu}\right|_{{\rm on-shell}}-\frac{1}{2}{\epsilon_{vu}}^{\eta\lambda}\left.C_{\eta\lambda vu}\right|_{{\rm on-shell}}=\left(2a'^2-a a''\right)\partial_w^2\phi/4=0 \ .
\ee
The only solution to this equation, without imposing any constraints on $\phi$, requires the scale factor to be either a constant or proportional to $1/\tau$. Thus, the on-shell self-duality equations can only be reduced to equations of motion of a scalar theory in flat \cite{Plebanski:1975wn} or (A)dS background \cite{Nagy:2022xxs}. Instead of constraining the scale factor, we could impose $\partial_w^2\phi=0$, but then the $vu\bar{w}u$ component of Eq.~\eqref{eq:onshell1} would lead to 
\be
\left.C_{vu\bar{w}u}\right|_{{\rm on-shell}}-\frac{1}{2}{\epsilon_{vu}}^{\eta\lambda}\left.C_{\eta\lambda \bar{w}u}\right|_{{\rm on-shell}}=\left(\partial_\tau-2a'/a\right)\left(2a'^2-a a''\right)\partial_w\phi/8=0 \ .
\ee
which again reduces the equations to the (A)dS and flat cases or further imposes $\partial_w\phi=0$. Imposing the latter would lead to a non-interacting theory, as can be seen in Eq.~\eqref{eq:ddphiZero} below.  Thus, to find a solution in more general backgrounds, we must consider the less restrictive constraint in \eqref{eq:SDgravity}, which we will refer to as off-shell self-duality. 

%%%%%%%%%%%%%%%%%%%%%%%%%%
\subsection{Off-shell self-duality} \label{sec:SDWeyl}

Following the result above, we analyze whether the off-shell self-duality constraint in ~\eqref{eq:SDgravity} can be solved in a similar manner for more generic backgrounds. Since this equation is invariant under Weyl rescalings of the metric, the scale factor will not play a role. In other words, every solution obtained in this section does not correspond to a single metric but a conformal class of metrics. 

We start by solving the components of Eq.~\eqref{eq:SDgravity} that are linear in $h_\mn$ subject to the ansatz in \eqref{eq:SDfrw} and \eqref{eq:hphi} and assumptions 1-3 described in the previous subsection. This requires that
\be \label{eq:sdLin}
  C_{vuvu}-\frac{1}{2} {\epsilon_{vu}}^{\eta \lambda} C_{\eta \lambda vu}\propto \partial^2_{w}h_{\bar{w}\bar{w}}-2\partial_u\partial_w h_{v\bar{w}}+\partial_u^2 h_{vv}=0 \ .
\ee
This equation can be solved by taking 
\begin{subequations}
\label{eq:Dops}
\eqs{
\Dop &=\Pi=\left(\partial_w,\partial_u\right)\ ,\label{flat_pies} \\
\tilde{\Dop} &=\Pi^\zeta\equiv \left(\partial_w,\partial_u+2  \,\zeta(u+v)\right) \ .
}
\end{subequations}
where $\zeta(u+v)$ is a function with units of inverse length. Using these operators, the self-dual equations reduce to only two independent equations, non-linear in $\phi$, given by the components $v,u,\bar{w},v$ and $\bar{w},u,\bar{w},u$ of Eq.~\eqref{eq:SDgravity}. The $v,u,\bar{w},v$ component determines the equation of motion satisfied by the scalar field:
\eqs{
&\left(-\partial_{u}\partial_{v}+\partial_{w}\partial_{\bar{w}}\right)\phi-\frac{\partial_u \zeta}{\zeta}\left(\partial_{u}+\partial_{v}\right)\phi-\frac{\partial^2_u \zeta}{\zeta}\phi \nonumber \\
&+\left(h_{vv}h_{\bar{w}\bar{w}}-h_{v\bar{w}}^{2}+\left(\frac{\partial^2_u \zeta}{2\zeta}-2\partial_u \zeta+\zeta^2\right)\left(\partial_{w}\phi\right)^{2}\right)=0. \label{eq:phieomf}
}
Then, the $\bar{w},u,\bar{w},u$ component of the self-dual equations can only be solved if
\be
\partial_u\left(\frac{\partial_u \zeta}{\zeta^2}\right)=0 \ , \label{eq:fEq}
\ee
which gives\footnote{Here and throughout the paper, we will ignore additional freedom corresponding to a constant shift of conformal time.} 
\be
\zeta=\frac{-2}{(u+v)}\ , \quad \text{or} \quad  \zeta=\text{constant} \ .
\ee
The details of this calculation can be found in the Appendix \ref{ap:sdGeneral}. The first case corresponds to the (A)dS self-dual equation. The constant $\zeta$ case will give rise to a new solution if the constant does not vanish and reduces to the flat one when $\zeta\rightarrow0$.

Using Eq.~\eqref{eq:fEq}, the equation of motion Eq.~\eqref{eq:phieomf} can be rewritten as
\be
\sqrt{g_\zeta}\left(\Box_{\zeta}-\frac{R_\zeta}{6}\right)\phi-\left\{ \left\{ L\zeta \ \phi,L\zeta \ \phi\right\} \right\} _{\zeta}=0 \ ,
\label{eq:SDeomF}
\ee
where $\Box_{\rm{\zeta}}$ is the Laplacian operator for an auxiliary metric $(g_{\mu\nu})_\zeta$ of the form \eqref{eq:SDfrw} with $h_{\mu \nu}=0$ and scale factor $a=L\zeta$. Note, however, that the scale factor of the auxiliary metric should not be identified with that of \eqref{eq:SDfrw} at this stage. Indeed, the off-shell self-duality in \eqref{eq:SDgravity} is Weyl invariant, so the scale factor in \eqref{eq:SDfrw} plays no role. We have introduced a length scale, $L$, to keep the scale factor dimensionless. Additionally, having the explicit factors of this length scale in the equations of motion will allow us to take the correct flat space limit. The auxiliary metric corresponds to de Sitter if $\zeta=-2/(u+v)$ or flat space if $\zeta=$constant. $R_\zeta$ is its Ricci scalar, and $g_\zeta$ its determinant. Note that the kinetic term is that of a conformally coupled scalar and can be mapped to a massless kinetic term by performing a Weyl transformation of the auxiliary metric to flat space. The interactions are given by a bracket defined as
\eqs{ \label{eq:JacobiZeta}
\left\{ f,g\right\} _{\zeta}&=\left\{ f,g\right\} +c_\zeta \,\zeta(u+v) \left(f\partial_{w}g-g\partial_wf\right) \ , \\
c_\zeta&\equiv2\left(\frac{\partial_u \zeta}{\zeta^2}-1\right)=\text{constant}    \ ,
}
where the undeformed Poisson bracket is defined in \eqref{poisson_def}. This bracket satisfies the Jacobi identity,
\be
\left\{f,\left\{ g, h\right\} _{\zeta}\right\} _{\zeta}+\left\{g,\left\{ h, f\right\} _{\zeta}\right\} _{\zeta}+\left\{h,\left\{ f, g\right\} _{\zeta}\right\} _{\zeta}=0.
\ee
and instead of satisfying the Leibniz rule, it satisfies a deformed version of it,
\be
\left\{ fg,h\right\} _{\zeta}=f\left\{ g,h\right\}_{\zeta}+g\left\{ f,h\right\}_{\zeta}-c_\zeta \zeta(u+v) \ fg\partial_{w}h \ ,
\ee
As shown in Appendix~\ref{ap:Jacobi}, this bracket corresponds to a Jacobi bracket \cite{JacobiVaisman,cabauBook}, which is defined as a Lie bracket on the algebra of smooth functions and is given by a bilinear first-order differential operator $D$ as
\be\label{Jac_brack_def}
D(f, g)=i(P)(\ud f \wedge \ud g)+f \ i(X) \ud g-g \ i(X) \ud f
\ee
where $i$ denotes the interior product, $P$ is a bivector, and $X$ a vector (called the Reeb vector field), satisfying
\be\label{conds_Jacobi}
[P, P]=2 X \wedge P \ , \quad [X, P]=0 \ ,
\ee
as a consequence of the Jacobi identity. Here, $[\ , \ ]$ is the Schouten-Nijenhuis bracket, which is a generalization of a Lie bracket for multivector fields. Using a coordinate basis, \eqref{Jac_brack_def}  can be expressed as 
\be\label{comp_Jacobi_bracket}
D(f, g)=P^{\mu\nu}\partial_\mu f \partial_\nu g +f X^\mu \partial_\mu g - g X^\mu \partial_\mu f \ ,
\ee
where $P^{\mu\nu}$ is anti-symmetric. Then \eqref{eq:JacobiZeta} can be put in the form above with $P^{\mu\nu}$ and $X^\mu$ given in \eqref{P_and_X_explicit}. It is interesting to note that other formulations of self-dual solutions on AdS backgrounds obtained from twistor space also give rise to a Jacobi instead of a Poisson bracket \cite{Bittleston:2024rqe,Taylor:2023ajd}.

The Poisson bracket, defined by Eq.~\eqref{eq:JacobiZeta} with $\zeta=0$, defines a kinematic algebra that can be lifted to the $w_{1+\infty}$ algebra \cite{Monteiro:2022lwm}. In the following, we derive the deformation of the $w_{1+\infty}$ algebra that arises when considering instead the Jacobi bracket with $\zeta\neq0$. By performing a Weyl rescaling $\phi\rightarrow (1/ (L\zeta))\phi$ one can rewrite Eq.~\eqref{eq:SDeomF} as
\be
\Box_{\mathbb{R}^{4}}\phi-\frac{1}{L\zeta} \left\{ \left\{ \phi,\phi\right\} \right\} _{\zeta}=0 \ , \label{eq:SDeomFlat}
\ee
The solutions to \eqref{eq:SDeomFlat} are plane waves: 
\be
\phi=e^{i k\cdot x} \ .
\ee
We will further consider on-shell states with $k^2=$0, and take the soft limit $(k_{u},k_{v},k_{w},k_{\bar{w}})\ \to\ (0,0,0,0)$ in such a way that $k_{\bar{w}}/k_{u}=k_{v}/k_{w}=\rho$, where $\rho$ is some number. These on-shell plane waves can now be written as an expansion in soft momenta given by
\be
e^{ik\cdot x}=\sum_{a,b=0}^{\infty}\frac{\left(ik_{u}\right)^{a}\left(ik_{w}\right)^{b}}{a!b!}\mathfrak{e}_{ab} \ ,
\ee
where $\mathfrak{e}_{ab}=\left(u+\rho \bar{w}\right)^{a}\left(w+ \rho v\right)^{b}$. Further defining $w_{m}^{p}=\frac{1}{2}\mathfrak{e}_{p-1+m,p-1-m}$ we find that
\eqs{
\left\{ w_{m}^{p},w_{n}^{q}\right\} _{\zeta}&=\left\{ w_{m}^{p},w_{n}^{q}\right\} +\frac{c_\zeta}{2} (m+q-p-n) \zeta(u+v) w_{m+n+1/2}^{p+q-3/2} \ .
}
When $\zeta=0$, this reduces to the $w_{1+\infty}$ algebra in flat space \cite{Fairlie:1990wv,Strominger:2021mtt}, and $\zeta\neq0$ gives a deformed version of this algebra.  There are known deformations of the $w_{1+\infty}$ algebra \cite{Pope:1991ig,Bittleston:2023bzp,Bu:2022iak,etingof2020newrealizationsdeformeddouble}, which involve a constant parameter. In principle, this is similar to the case $\zeta=\text{constant}\neq0$, but it is unclear whether our case corresponds to any of the known deformations via a change of variables.

%%%%%%%%%%%%%%%%%%%%%%%
\section{Color-kinematics duality and double copy} \label{sec:DoubleCopy}
An intriguing property of self-dual gravity in a flat background is that it can be derived from the double copy of self-dual Yang-Mills at the Lagrangian level \cite{Monteiro:2011pc}. Since Yang-Mills theory is classically scale-invariant in four dimensions, its Lagrangian in any conformally flat background is the same as in flat space, although one has to impose boundary conditions if there is a boundary. Moreover, self-dual gravity in AdS can be derived from an asymmetrical double copy \cite{Lipstein:2023pih}. We will briefly review the self-dual Yang-Mills theory and then show how to obtain self-dual gravity in more general backgrounds from a double copy.

\subsection{Self-dual Yang-Mills}
We begin by summarizing the construction of the self-dual solution for gauge theories. In this case, the self-dual condition is given as
\be
F_{\mu\nu}=\frac{1}{2}\epsilon_{\mu\nu\rho\lambda}F^{\rho\lambda} \ ,
\ee
where $F_{\mu\nu}$ is the YM fields strength. Due to scale invariance, this equations imposes the same constraint in any conformally flat background. Working in lightcone gauge, $A_{u}=0$, the solution is given by
\be
\label{eq:sdYM}
A_i=0, \quad A_\alpha=\Pi_\alpha \Phi .
\ee
with $\Pi_\alpha$ as in \eqref{flat_pies} and where $\Phi$ is a scalar field in the adjoint representation of the gauge group whose equation of motion is
\eqs{
\Box_{\mathbb{R}^{4}}\Phi-i[\{\Phi,\Phi\}]=0 \  \label{eq:sdYMflat}, \\
[\{f,g\}]= \varepsilon^{\alpha\beta} \left[\Pi_\alpha f,\Pi_\beta g\right] \ ,
}
and  $[ \ ,\ ]$ is the standard Lie bracket of the gauge theory. While the same scalar field theory describes the self-dual Yang-Mills solutions in all conformally flat spacetimes, the boundary conditions will be different since Weyl transformations change the nature of the asymptotic structure of the spacetime. For example, the lack of time translations in FLRW will be explicit when calculating boundary correlation functions. Eq.~\eqref{eq:sdYMflat} is solved by plane waves
\be
\Phi=c e^{i k\cdot x} \ , 
\ee
where $c$ is a spacetime constant in the adjoint representation, $k\cdot x$ is the flat space inner product, and $k$ satisfies the on-shell condition $k_{u}k_{v}-k_{w}k_{\bar{w}}=0$. Using these solutions as external states, the three-point vertex is given by 
\eqs{
V_{\rm{SDYM}} &=\frac{1}{2} \, X\left(k_{1},k_{2}\right)f^{a_{1}a_{2}a_{3}}, \\
X\left(k_{1},k_{2}\right)&=k_{1u}k_{2w}-k_{1w}k_{2u}
}
where $f^{a_{1}a_{2}a_{3}}$ are the structure constants of the color algebra and the factor $X\left(k_{1},k_{2}\right)$ can be thought of as the structure constants of the kinematic algebra which in this case corresponds to area-preserving diffeomorphisms in the $u-w$ plane. It is also worth highlighting that the Jacobi identity for $X\left(k_{1},k_{2}\right)$ is satisfied for off-shell momenta, that is, without imposing $k^2=0$. This displays one of the simplest realizations of color-kinematics duality. 

\subsection{Double copy}
The self-dual Yang-Mills and gravity solutions for conformally flat spacetimes that we have described above are given in terms of a scalar field satisfying
\begin{align}
\Box_{\mathbb{R}^{4}}\Phi-i[\{\Phi,\Phi\}]=0 \ , \label{sdymeq1} \\
\Box_{\mathbb{R}^{4}}\phi-\frac{1}{L\zeta} \left\{ \left\{ \phi,\phi\right\} \right\} _{\zeta}=0  \ , \label{sdgeq1}
\end{align}
respectively. Given the explicit color-kinematics duality in the self-dual Yang-Mills theory, one can obtain straightforwardly the double copy by exchanging color by kinematics. At the level of the equations of motion, one can perform the replacements
\be
\Phi \to \phi\ ,\quad \quad i[\quad] \to  \frac{1}{L \zeta} \{\quad\}_{\zeta} \label{eq:DCeom}
\ee
to go from the self-dual Yang-Mills equation in \eqref{sdymeq1}
to the self-dual gravity one in \eqref{sdgeq1}. This gives an asymmetric double copy since the gravitational equation of motion, Eq.~\eqref{eq:SDeomFlat}, involves both the flat self-dual Poisson bracket and the cosmological Jacobi bracket. Note that throughout the paper, we have set the couplings of both Yang-Mills and gravity to one, which is why we do not have the usual replacement $g\rightarrow\kappa$. While this might look singular in the flat space limit, $\zeta\rightarrow 0$, we can recover the flat self-dual equations of motion by keeping $L \zeta$ fixed and taking $\zeta\rightarrow 0$ as explained in Sec.~\ref{sec:SDWeyl}. 

Similarly to the self-dual Yang-Mills case, the self-dual gravity equation in \eqref{sdgeq1} allows for plane wave external states. The Feynman rule for the three-point vertex for such states is then given by
\eqs{
V_{\rm{SDG}}&=\frac{1}{2} \,\frac{1}{L \zeta} \, X\left(k_{1},k_{2}\right) X^\zeta\left(k_{1},k_{2}\right) \ , \\
X^\zeta\left(k_{1},k_{2}\right)&=X\left(k_{1},k_{2}\right)-i \, c_\zeta \, \zeta \left(k_{1}-k_{2}\right)_{w} \ .
}
Thus, the double copy replacement for the three-point vertex is 
\be
f^{a_{1}a_{2}a_{3}}\rightarrow \frac{1}{L \zeta}{X^\zeta}\left(k_{1},k_{2}\right) \ . \label{eq:DCvertex}
\ee
As before, the flat space limit is taken by keeping $L \zeta$ fixed and taking $\zeta\rightarrow 0$.

We have formulated the double copy by using the equations of motion of both Yang-Mills and gravity with a flat auxiliary metric. We could have equivalently performed a field redefinition of the scalar fields, $\Phi\rightarrow a\Phi$ and $\phi\rightarrow a\phi$, which changes the kinetic term to a conformally coupled scalar in an FLRW background with scale factor $a$.  Under these rescalings, the equations of motion read
\begin{align}
\sqrt{g_{a}}\left(\Box_{a}-\frac{R_{a}}{6}\right)\Phi-i \, a[\{a\,\Phi,a\,\Phi\}]=0 \ , \label{eq:SDYMcurved} \\
\sqrt{g_a}\left(\Box_{a}-\frac{R_a}{6}\right)\phi-\frac{a}{L\zeta} \left\{ \left\{ a \ \phi,a \ \phi\right\} \right\} _{\zeta}=0 \ . \label{eq:SDcosmocurved}
\end{align}
The double copy is again given by the color-kinematic replacements in Eq.~\eqref{eq:DCeom} and Eq.~\eqref{eq:DCvertex}. Note that this is not a different double copy in a new background, but simply a field redefinition of the conformally coupled scalars. The double copy that we have formulated here is a double copy for all conformal classes, not a double copy on a specific FLRW background. 

%%%%%%%%%%%%%%%%%%%%
\section{Cosmological self-dual solutions} \label{sec:sdCosmology}
In the previous section, we derived general solutions to the off-shell self-duality equation in \eqref{eq:SDgravity}. Demanding that they obey the on-shell self-duality equation in \eqref{onshellsd} where both the Ricci tensor and the Ricci scalar are fixed by the Einstein equations then fixes the background to be either flat or AdS. In this section, we will only impose a constraint on the Ricci scalar, leaving the traceless part of the Ricci tensor unconstrained. Since we focus on cosmological solutions, we will require the Ricci scalar, or equivalently, the trace of the stress-energy tensor to be that of an FLRW metric. In the present case, the self-dual solutions above have
\eqs{
R^\text{s.d.}&=R^\text{FLRW}-\frac{24(\zeta\partial_ua-\partial_u^2a)\partial_w^2\phi}{a^3}\ , \\
R^\text{FLRW}&=\frac{24\partial_u^2a}{a^3} \ ,
}
so that imposing 
\be
R^\text{s.d.}=R^\text{FLRW} \ ,
\ee
leads to
\be
\zeta(u+v)\equiv\frac{\partial_u^2a}{\partial_u a}=\frac{\partial_uH}{H}+aH \  , \label{eq:FixR}
\ee
where $H$ is the Hubble parameter. The usual expression for the Hubble parameter is given in terms of cosmic time, defined by $\ud t= a \ud \tau $, so that $H=\partial_ta/a$. In terms of lightcone coordinates, we have $H=2(\partial_u a)/a^2$. The requirement of a self-dual Weyl tensor in Eq.~\eqref{eq:fEq} implies that we need to take either 
\be
a=e^{c \tau/L} \ , \ \text{with $c=$ constant} \quad  \text{or} \quad a=(\tau/L)^p \ , \ \text{with} \ \ p=0,-1,1 \ ,
\ee
where the length scale $L$ determines the curvature of the spacetime. In the exponential case, this is the scale factor of a coasting FLRW spacetime. For the power law, the first two cases correspond to the flat and (A)dS solutions described above, while the third one gives the scale factor of a radiation-dominated Universe. We explore the new cases, radiation-dominated and coasting FLRW, in more detail in the following sections. 

The source of these cosmological self-dual metrics can be interpreted as a viscous fluid with a stress-energy tensor given by
\be
T^\mn=\rho u^\mu u^\nu+ P \gamma^\mn- 2 \eta \sigma^\mn + q^{(\mu}u^{\nu)} \ , \label{eq:TmunuVisc}
\ee
where $\rho,P,u^\mu,$ and $\gamma^\mn$ are defined under Eq.~\eqref{eq:TmunuPerfectFluid}, $\eta$ is the shear viscosity , $q^\mu$ is the momentum density, and the traceless tensor, $\sigma^\mn$ is the shear tensor (or anisotropic stress perturbation) \cite{Andersson:2006nr}. The equation of state parameter is defined as 
\be
\omega=P/\rho  \ .
\ee
Below, we will examine the properties of the sources for the different cosmological self-dual solutions.

One should remember that we are obtaining a non-perturbative result, but it can be helpful for those familiar with cosmology to understand $h_\mn$ as the perturbations that would arise in standard cosmological perturbation theory \cite{Ma:1995ey}. In that case, these perturbations are sourced by the deviations from the perfect fluid sources. In the present case, one can choose appropriate boundary conditions on the scalar field and match any desired physical boundary conditions. For example, the boundary conditions can be chosen to have an asymptotically FLRW spacetime \cite{Bonga:2020fhx,Enriquez-Rojo:2022onp}.

We have previously formulated a general double copy prescription for all the conformal classes of cosmological self-dual metrics. We can restrict this procedure to the four special solutions with $R^\text{s.d.}=R^\text{FLRW}$, and consider now a double copy on a fixed FLRW background. The color-kinematic replacements for the equations of motion and three-point vertex can be found in Table~\ref{tab:ckEOM} and Table~\ref{tab:ckVertex} respectively.

\begin{table}[!ht]
    \centering
    \begin{tabular}{|c|c|c|}
  \hline
  SD solution & scale factor & color-kinematics replacements \\
  \hline
  Flat & $a=1$ & $ \quad \Phi \to  \phi, \quad \ i[\quad] \to  \phantom{\frac{1}{a}} \{\quad\}\phantom{_{\zeta=\mathcal{H}/2}}$ \\
  \hline
  Radiation & $a=\frac{\tau}{L}$ & $ \quad \Phi \to  \phi, \quad \ i[\quad] \to  \phantom{\frac{1}{a}} \{\quad\}\phantom{_{\zeta=\mathcal{H}/2}}$ \\
  \hline
  (A)dS & $a=\frac{-2L}{\tau}$ & $ \quad \Phi \to \phi, \quad \ i[\quad] \to  \frac{1}{a} \{\quad\}_{\zeta=a}\phantom{_{/2}}$ \\
  \hline
  Non-acc. FLRW & $a=e^{\mathcal{H}\tau}$ & $\quad  \Phi \to \phi, \quad \ i[\quad] \to  \phantom{\frac{1}{a}}\{\quad\}_{_{\zeta=\mathcal{H}/2}}$ \\
  \hline
\end{tabular}
    \caption{Color-kinematics replacements for self-dual gravity in different backgrounds. The replacements should be applied to the self-dual Yang-Mills solution as given by Eq. \eqref{eq:SDYMcurved}.}
    \label{tab:ckEOM}
\end{table}

\begin{table}[!ht]
    \centering
    \begin{tabular}{|c|c|c|}
  \hline
  SD solution & scale factor &color-kinematics replacements \\
  \hline
  Flat & $a=1$ & $ f^{a_{1}a_{2}a_{3}}\rightarrow {X}\left(k_{1},k_{2}\right),$ \\
  \hline
  Radiation & $a=\frac{\tau}{L}$ & $ f^{a_{1}a_{2}a_{3}}\rightarrow {X}\left(k_{1},k_{2}\right),$ \\
  \hline
  (A)dS & $a=\frac{-2L}{\tau}$ & $ f^{a_{1}a_{2}a_{3}}\rightarrow\frac{1}{a}X^{\zeta=a}\left(k_{1},k_{2}\right),$ \\
  \hline
  Non-acc. FLRW & $a=e^{\mathcal{H}\tau}$ & $ f^{a_{1}a_{2}a_{3}}\rightarrow X^{\zeta=\frac{\mathcal{H}}{2}}\left(k_{1},k_{2}\right),$ \\
  \hline
\end{tabular}
    \caption{Color-kinematics replacements for the three-point vertices of self-dual gravity in different backgrounds.}
    \label{tab:ckVertex}
\end{table}

%%%%%%%%%%%%
\subsection{Self-dual radiation-dominated FLRW} \label{sec:radiation}
This section analyzes self-dual gravitational solutions sourced by a traceless stress-energy tensor with a time-dependent scale factor. We dub these solutions as self-dual radiation. As mentioned in the previous section, we can find a self-dual solution with a metric 
\be
ds^2=\left(\frac{u+v}{2 L}\right)^2 \left(dw\,d\bar{w}-du\,dv +\frac{1}{4}\left(\Pi_{(\alpha}{\Pi^\zeta}_{\beta)}\phi \right)\,dx^\alpha dx^\beta\right) \ .
\ee
When $\phi=0$, this reduces to the radiation-dominated FLRW solution. From \eqref{eq:FixR}, we see that $\zeta=0$ for $a=\tau/L$. Thus the scalar satisfies the same equation as it does in flat background, \eqref{scalareomflat}. After a field redefinition that takes $\phi\rightarrow a\phi$, the equation of motion can be rewritten in terms of the Laplacian for the FLRW radiation metric as
\be
\sqrt{g_a} \Box_{a} \phi -  a\{\{a\phi,a\phi\}\}=0 \ ,
\ee
where $a=(u+v)/2L$ and the conformally coupled mass term vanishes since the Ricci scalar vanishes in this background.

The energy density and pressure of the source of the self-dual radiation solution are given by
\eqs{
\rho=\frac{M_{Pl}^2}{L^2} & \left( \frac{3+3\partial_w^2\phi+(\partial_w^2\phi)^2}{a^4}  -\frac{L(\partial_u+\partial_v+2(\partial_u\partial_w\phi)\partial_w)\partial^2_w\phi}{a^3} \right)\ , \\
P=\frac{1}{3} \rho \ .
}
Both redshift as expected for a radiation component as long as $\partial^2_w\phi\ll1$. Similarly, one can find that the trace of the stress-energy tensor vanishes, ${T_\mu}^\mu=0$, or equivalently that the equation of state is $\omega=1/3$, which identifies the source as radiation. Note that contrary to the standard perfect fluid sources of FLRW spacetimes, the source of this metric does not have to be homogeneous or isotropic.  We provide the full expression for the stress-energy tensor in an ancillary Mathematica file. 

%%%%%%%%%%%%%%%%%%%%
\subsection{Self-dual coasting FLRW} \label{sec:nonAccel}
An FLRW Universe with $a(\tau)=e^{\mathcal{H}\tau}$ and $\mathcal{H}\equiv\partial_\tau a/a=\text{constant}$ is sourced by a perfect fluid with an equation of state $\omega=-1/3$. Going back to Cartesian coordinates and performing a further change of coordinates to cosmic time, $a(\tau)\ud \tau=\ud t$, we write the background metric in its better-known form
\be
\ud s^2=-\ud t^2+(\mathcal{H}t)^2\ud\mathbf{x}^2 \ ,
\ee
which describes a coasting FLRW cosmology \cite{Kolb:1989bg}. When written in this form, one can describe whether the Universe is accelerating by looking at the dimensionless parameter
\be
q=-a\frac{\partial^2_ta}{(\partial_ta)^2} \ ,
\ee
which is referred to as the deceleration parameter. The Universe is accelerating for $q<0$, decelerating for $q>0$, and neither if $q=0$, as in the metric above.

We have found a self-dual solution with the Ricci-scalar of this coasting FLRW spacetime. This solution has a metric of the form
\be
ds^2=\left(e^{\mathcal{H}\frac{u+v}{2}}\right)^2 \left(dw\,d\bar{w}-du\,dv +\frac{1}{4}\left(\Pi_{(\alpha}{\Pi^\zeta}_{\beta)}\phi \right)\,dx^\alpha dx^\beta\right) \ .
\ee
Noticing that the function $\zeta$ in Eq.~\eqref{eq:FixR} can be written in terms of the deceleration parameter as 
\be
\zeta=(1-q)\mathcal{H}/2 \ ,
\ee
we see that in the present case $\zeta=\mathcal{H}/2$. From \eqref{eq:phieomf}, we then find that the scalar field satisfies the equation of motion 
\be
\left(\partial_{u}\partial_{v}-\partial_{w}\partial_{\bar{w}}\right)\phi+\left(h_{vv}h_{\bar{w}\bar{w}}-h_{v\bar{w}}^{2}+(\mathcal{H}/2)^2
\left(\partial_{w}\phi\right)^{2}\right)=0 \ ,
\ee
which under a field redefinition, $\phi\rightarrow a\phi$ can be rewritten as
\be
\sqrt{g}(\Box_{a}+(\mathcal{H}/a)^2)\phi -   a\{\{a\phi,a\phi\}\}_{\zeta=\mathcal{H}/2}=0 \ ,
\ee
where $a=e^{H\tau}$, and the mass term is simply the conformal coupling in the coasting FLRW spacetime.

We can compute the source for this solution and find that it is a viscous fluid with energy density
\be
\rho=\frac{(M_{Pl}\mathcal{H})^2}{a^2}\left(3+\hat{\Theta}_\text{exp.}\partial_w^2\phi\right) \ ,
\ee
where $\hat{\Theta}_\text{exp.}$ is a differential operator given by 
\be
\hat{\Theta}_\text{exp.}=1+\mathcal{H}^{-1}\partial_w\partial_u\phi\partial_w \ ,
\ee
and the equation of state of the fluid is
\be \label{eq:EoSnona}
\omega=-\frac{1}{3}+\frac{\hat{\Theta}_\text{exp.}\partial_w^2\phi}{6+\hat{\Theta}_\text{exp.}\partial_w^2\phi} \ . 
\ee
For $\partial_w^2\phi\ll1 $, this approaches the usual equation of state of a coasting FLRW spacetime. In other words, it will have a slight acceleration or deceleration depending on the sign of the second term in Eq.~\eqref{eq:EoSnona}. The full expression for the stress-energy tensor is found in the ancillary Mathematica file. 

%%%%%%%%%%%%%%%%%%%%
\subsection{Approximately FLRW self-dual metrics} 
\label{sec:ConformalSD}
While there are only four cases of self-dual cosmological solution with a Ricci scalar fixed to be that of an FLRW metric, there are an infinite set of metrics with self-dual Weyl tensor which fall into three conformal classes mentioned in Section \ref{sec:SDWeyl}. Note that the self-dual solution in radiation domination background can be obtained by performing a Weyl transformation of the one in flat background so it belongs the same conformal class. Hence, the three conformal classes can be obtained by performing Weyl transformations of the flat, dS, and coasting solutions constructed in the previous section. In this section we will construct other representatives of these conformal classes whose stress-energy tensor approximately behaves as an FLRW one, as long as the scalar field satisfies a simple additional constraint.

Given a solution to the off-shell self-duality equation in \eqref{eq:SDgravity}, one can obtain another solution by performing a Weyl transformation:
\begin{equation}
    \tilde{g}_{\mu\nu}=\Omega^2(\tau) g_{\mu\nu} \ . \label{eq:Conf}
\end{equation}
The new metric $\tilde{g}_{\mu\nu}$ is sourced by a stress-energy tensor given by
\eqs{
 \frac{ \tilde{T}_{\mu\nu}}{M_{Pl}^2}=&\frac{T_{\mu\nu}}{M_{Pl}^2}-2\frac{\nabla_\mu\nabla_\nu\Omega}{\Omega}+4\frac{\nabla_\mu\Omega\nabla_\nu\Omega}{\Omega^2} \nonumber  \\
    &-g_{\mu\nu}\left(-2\frac{\nabla_\mu\nabla^\mu\Omega}{\Omega}+4\frac{\nabla_\mu\Omega\nabla^\mu\Omega}{\Omega^2}\right),
}
where $T_{\mu\nu}$ is the stress-energy tensor sourcing $g_{\mu\nu}$ and $\nabla_\mu$ is its covariant derivative. The trace of the stress-energy tensor is
\begin{equation}
\tilde{T}=\frac{1}{\Omega^2}\left(6M_{Pl}^2 \frac{\nabla^2 \Omega}{\Omega}+T\right) \ ,
\end{equation}
where $T={T_\mu}^\mu$. Since this new metric is no longer a homogeneous isotropic metric, the metric is not expected to be sourced by a perfect fluid but rather by a viscous fluid with a non-zero momentum flux vector and shear tensor, as in Eq.~\eqref{eq:TmunuVisc}. The energy density  and pressure of the fluid are
\eqs{
    \tilde{\rho}=\tilde{T}_{\mu\nu}\tilde{u}^\mu\tilde{u}^\nu \ , \quad
    \tilde{P}=\frac{1}{3}\tilde{T}_{\mu\nu}\tilde{\gamma}^{\mu\nu} \ 
}
where $\tilde{u}^\mu$ is a unit timelike vector with respect to $\tilde{g}_{\mu\nu}$ and the metric of the surface perpendicular to $u^\mu$ is $\tilde{\gamma}_{\mu\nu}=\tilde{g}_{\mu\nu}+\tilde{u}_\mu\tilde{u}_\nu$. 

We will now perform Weyl transformations of self-dual solutions in flat, dS, and non-accelerated FRLW backgrounds such that the resulting metric takes the following general form:
\begin{equation}
\ud s^2=\left(\frac{u+v}{2 L}\right)^{2p}\left(-\ud u \ud v+\ud w\ud\bar{w}+\frac{1}{4}\left(\Pi_{(\alpha}{\Pi^\zeta}_{\beta)}\phi \right)\,dx^\alpha dx^\beta\right)\ , \label{eq:confSD}
\end{equation}
which describes well-known power law cosmologies when $\phi=0$, see Appendix \ref{ap:Cosmo}. This metric can be obtained by applying a Weyl transformation with $\Omega=(\tau/L)^p$ to the solution in flat background, $\Omega=(\tau/L)^{p+1}$ to the one in dS background, and $\Omega=e^{-\mathcal{H}\tau}(\tau/L)^p$ to the one in coasting background. The properties of the resulting stress-energy tensors are described in detail in Appendix \ref{ap:ConfSD}, and their energy density and equation of state take the schematic form
\begin{align*}
\rho&=\rho_\text{FLRW}+\hat{\Theta}_\zeta \ \partial^2_w\phi \ , \\
\omega&=\omega_\text{FLRW}+\hat{\Gamma}_\zeta \ \partial^2_w\phi \ ,
\end{align*}
where $\rho_\text{FLRW}$ and $\omega_\text{FLRW}$ are the energy density and equation of state of the source of the metric in Eq.~\eqref{eq:confSD} with $\phi=0$, and $\hat{\Theta}_\zeta$ and $\hat{\Gamma}_\zeta$ are differential operators depending on the conformal class we are working with. Note that if we require that $\partial^2_w\phi\ll1$, these solutions have sources that approximate those of the corresponding FLRW metric. This case is closer to what happens in cosmological perturbation theory; the equation of state remains close to the FLRW one, but it is not forced to remain the same. Strictly imposing that $\partial^2_w\phi=0$, the scalar can be written as 
\be
\phi=\phi_1(u,v,\bar{w})+w \phi_2(u,v,\bar{w}) \ ,
\ee
where the equation of motion now reduces to
\eqs{
\sqrt{g_a}\left(\Box_{a}-\frac{R_a}{6}\right)\phi_1+a^2\partial_{\bar{w}}\phi_2+\frac{a}{\zeta}\left(\partial_u(a\phi_2)\right)^2-c_\zeta \Phi_2\partial_u(a\phi_2)=0 \ , \label{eq:ddphiZero} \\
\sqrt{g_a}\left(\Box_{a}-\frac{R_a}{6}\right)\phi_2=0 \ .
}
We can see that $\phi_2$ is a free scalar whose $u$ and $\bar{w}$ derivatives source the scalar $\phi_1$.

As commented above, if we start with the self-dual solution in flat background and apply a Weyl transformation with $\Omega=\tau/L$, this gives the self-dual solution in radiation-dominated background. Similarly, if we Weyl transform the dS or non-accelerated self-dual solutions to flat background (such that the resulting metric takes the form in \eqref{eq:confSD} with $p=0$), the resulting stress tensor turns out to be traceless:  
\be
\tilde{T}\propto6M_{Pl}^2 a\nabla^2(1/a)+T=M_{Pl}^2 R+T=0 \ , 
\ee
where $R$ is the Ricci scalar of $g_\mn$, and we imposed the Eisntein equations in the last equality. On the other hand, the energy density does not evolve as the usual radiation-dominated FLRW ($\rho\sim \tau^{-4}$). Schematically, if we start with the dS self-dual solution and Weyl transform to flat background the resulting stress-energy tensor has $\rho\sim \tau^{-2}+\hat{\Theta}_\zeta\partial^2_w\phi$, and if we start with the coasting self-dual solution and Weyl transform to flat background we find that $\rho=\text{constant}+\hat{\Theta}_\zeta\partial^2_w\phi$. The exact expressions are given in Appendix \ref{ap:ConfSD}. 

%%%%%%%%

%%%%%%%%%%%%%%%%%%%%%
\section{Conclusions and discussion} \label{sec:Concl}
In this paper, we show that an infinite set of metrics with a self-dual Weyl tensor can be described using conformally coupled scalars with cubic interactions containings Jacobi brackets. These metrics look like a time-dependent deformation of the well-known solution for self-dual gravity in flat space. In particular we find three distinct conformal classes of self-dual metrics: flat, (A)dS, and coasting FLRW. We also present a general double copy prescription that maps self-dual Yang-Mills in an FLRW background to these self-dual cosmological solutions and show that they exhibit a deformed $w_{1+\infty}$ algebra, generalising the one found for AdS in \cite{Lipstein:2023pih}.

Interestingly, if we demand that the Ricci scalar of these self-dual solutions is equal to that of an FLRW metric, there are only four possible backgrounds for which this is possible: flat, dS, radiation-domination, and coasting FLRW. More general solutions can then be obtained by performing Weyl transformations of these solutions. While the solution corresponding to radiation domination can be obtained from a Weyl transformation of the flat solution, in general performing such Weyl transformations will lead to solutions whose Ricci tensor is not that of an FLRW metric. On the other hand, we find that the stress tensor of the resulting solutions corresponds to viscous fluids whose equations of state become FLRW-like in the limit $\partial_w^2\phi\ll1$. Interestingly, we find that Weyl-transforming the dS and non-accelerated self-dual solutions to the flat background results in a traceless stress tensor, which, therefore, describes a fluid whose equation of state is that of radiation.  

The existence of self-dual cosmological solutions and their intriguing Jacobi brackets suggests many future directions. An immediate question is how the color-kinematics duality at the level of equations of motion translates to correlation functions in FLRW spacetimes. While all tree-level amplitudes of self-dual Yang-Mills and self-dual gravity vanish beyond three points, this will not be the case for the curved background we consider because they are time dependent so energy is not conserved. Nevertheless, we expect the correlation functions in these backgrounds to be strongly constrained by symmetry. Finding an underlying geometric interpretation of the Jacobi brackets and recovering an infinite hierarchy of asymptotic symmetries, along the lines of \cite{Campiglia:2021srh}, would also be another important direction. Finally, it would be interesting to consider Moyal deformations of the scalar theories discussed in this paper. In a flat background, such deformations give rise to chiral higher spin theories \cite{Monteiro:2022xwq}, so doing so in the present context may give higher spin theories in cosmological backgrounds, which may be of interest for holography \cite{Jain:2024bza,Aharony:2024nqs}. 

\section*{Acknowledgments}
MCG work is supported by the Imperial College Research Fellowship. A.L. and S.N. are supported by an STFC Consolidated Grant ST/T000708/1.

%%%%%%%%%%%%%%%%%%%%%
\appendix

\section{Power law Cosmologies} \label{ap:Cosmo}
Power law cosmologies have the following metric:
\be
\ud s^2=\left(\frac{u+v}{2 L}\right)^{2p}\left(-\ud u \ud v+\ud w\ud\bar{w}\right)\ ,
\ee
where
\be
a(u+v)=\left(\frac{u+v}{2 L}\right)^{p} \ ,
\ee
and are sourced by a perfect fluid with stress-energy tensor 
\be
T^\mn=(\rho+P) u^\mu u^\nu+ P g^\mn \ ,
\ee
with an equation of state parameter \be
\omega=\frac{2-p}{3p} \ .
\ee
The metric, in this coordinates, is a decelerating (for $p>0$) or accelerating (for $p<0$) FLRW spacetime. We approach the flat space limit as $|u+v|\rightarrow \infty$ in both cases. In the decelerating case, the spacetime has a null infinity and its Penrose diagram is the upper half of the Minkowski one with a singularity at $u+v\rightarrow0$ corresponding to the Big Bang. These spacetimes include the matter-domination with $p=2$ and radiation-domination with $p=1$. On the other hand, the accelerating FLRW spacetimes have a spatial boundary at infinity, just like de Sitter. Taking $u+v\rightarrow -\infty$ now brings us to the far past; hence, the spacetime looks flat. Another relevant case, not included above, has a power law scale factor in cosmic time $t$, defined from $a(\tau)\ud \tau=\ud t$, but becomes exponential in conformal time. This is the case of an FLRW spacetime with no acceleration and $\omega=-1/3$. Its Penrose diagram is the same as the Minkowski one. Details on the conformal structure of these spacetimes can be found in \cite{Harada:2018ikn}.

%%%%%%%%%%%%%%
\section{Self-dual off-shell Weyl tensor} 
\label{ap:sdGeneral}
In this Appendix, we show how to solve the two independent self-dual equations that are non-linear in the scalar field. We start with the simplest one, which can be written as
\be
a^{-2}\left( C_{vu\bar{w}v}-\frac{1}{2} {\epsilon_{vu}}^{\eta \lambda} C_{\eta \lambda \bar{w}v} \right)=\frac{1}{2}\zeta(u+v)\ \partial_w \text{eom} \ ,
\ee
where eom is given by Eq.~\eqref{eq:phieomf}. The last self-dual equation is
\be \label{eq:sdLast}
\Psi=a^{-2}\left( C_{\bar{w}v\bar{w}v}-\frac{1}{2} {\epsilon_{\bar{w}v}}^{\eta \lambda} C_{\eta \lambda \bar{w}v} \right)\ .
\ee
Since we have fixed the equation of motion of $\phi$, to obtain a self-dual solution, we require that Eq.~\eqref{eq:sdLast} can be written entirely in terms of the equation of motion and its derivatives. When $\zeta=0$ we have that Eq.~\eqref{eq:sdLast} is
\begin{align*}
\Psi\big|_{\zeta=0}=(\partial_w\partial_{\bar{w}}-\partial_u\partial_v)(\text{eom})-\partial_u^2\phi\partial_w^2(\text{eom}) \nonumber \\ -\partial_w^2\phi\partial_u^2(\text{eom})+2\partial_w\partial_u\phi\partial_w\partial_u(\text{eom}) \ . \label{eq:PsiZeta0}
\end{align*}
To find a solution for a general scale factor we look at each derivative contribution separately:
\eqs{
\Psi=&\Psi_0+\zeta \Psi_1+ (\zeta^2 \Psi_2 + \zeta' \Psi_3) \nonumber \\
&+(\zeta^3 \Psi_4+\zeta\zeta' \Psi_5+\zeta''\Psi_6)\nonumber \\
&+(\zeta^2\zeta'\Psi_7+\zeta\zeta''\Psi_8+(\zeta')^2\Psi_9+\zeta''' \Psi_{10}) \ ,
}
where $
'\equiv\partial_{u+v}$. Writing $\Psi_0=\Psi\big|_{\zeta=0}$ with the equation of motion for $\zeta\neq0$ introduces new terms proportional to $\zeta$. We can now find an expression for $\Psi_1$ in terms of the equation of motion, which will also introduce new terms proportional to $\zeta$, but this won't affect the previous solutions since the new terms arise at a higher mass dimension. Thus, we can solve order by order in the mass dimension of the terms with $\zeta$ to find a complete solution. Proceeding this way, we find that
\eqs{
\Psi_1=&-2\partial_v (\text{eom})-2 \partial_u\phi \partial_w^2(\text{eom})+2\partial_u\partial_w\phi \partial_w (\text{eom}) \nonumber\\
&-2\partial_w^2\phi\partial_u(\text{eom})+2\partial_w \phi\partial_u\partial_w(\text{eom})\nonumber \\
&+\frac{\zeta'}{\zeta^2}(\partial_u (\text{eom})+\partial_v (\text{eom})) \ ,
}
But the solution breaks at the next order where there is no expression for $\Psi_2$ and $\Psi_3$ in terms of the equation of motion for a general $\zeta$. The obstruction arises due to the following term
\begin{equation*}
\Psi^\text{eom}\supset -\zeta\partial_u\left(\frac{\partial_u \zeta}{\zeta^2}\right)\left(\partial_u^2\phi+\partial_v^2\phi\right)  \ ,
\end{equation*}
which cannot be written in terms of the equation of motion and necessarily appears when rewriting lower-order contributions in terms of the equation of motion. To find a solution, we need to fix $\zeta$ so that this term vanishes. This is the case for
\be
\zeta=\frac{-2}{(u+v)}\ , \quad \text{or} \quad  \zeta=1/L \ ,
\ee
where $L$ is a constant length scale. As mentioned earlier, when $\zeta=0$, $\Psi$ is given by Eq.~\eqref{eq:PsiZeta0}. When $\zeta=\frac{-2}{(u+v)}$ 
\eqs{
&\Psi^{\zeta=\frac{-2}{(u+v)}}=\Psi\big|_{\zeta=0} +\zeta \Psi_1 +\frac{\zeta'''}{\zeta'}\left(\partial_w\phi\partial_w(\text{eom})\right)\nonumber \\
&+\frac{\zeta'^2}{\zeta^2}\left(6\partial_w\phi\partial_w(\text{eom})+2\phi\partial_w^2(\text{eom})+(6\partial_w^2\phi-1/2)(\text{eom})\right) \nonumber \\
&-\frac{\zeta''}{\zeta}\left(5\partial_w\phi\partial_w(\text{eom})+2\phi\partial_w^2(\text{eom})+
5\partial_w^2\phi(\text{eom})\right) \ .
}
Meanwhile, choosing $\zeta=1/L\neq0$ we find
\eqs{
&\Psi^{\zeta=c}=\Psi\big|_{\zeta=0} +\zeta \Psi_1 +\zeta^2\Big(\left(\partial_w\phi\partial_w(\text{eom})\right)\nonumber \\
&+6\partial_w\phi\partial_w(\text{eom})+2\phi\partial_w^2(\text{eom})+6\partial_w^2\phi(\text{eom})\nonumber \\
&-5\partial_w\phi\partial_w(\text{eom})-2\phi\partial_w^2(\text{eom})-
6\partial_w^2\phi(\text{eom})\Big) \ .
}

%%%%%%%%%%%%%%%%
\section{Jacobi bracket of cosmological self-dual solutions} \label{ap:Jacobi}
The bracket in \eqref{eq:JacobiZeta} corresponds to a Jacobi bracket. In this Appendix, we will explicitly show that this is the case. For convenience, we reproduce the component-wise definition of the Jacobi bracket in \eqref{comp_Jacobi_bracket} below:
\be\label{comp_Jacobi_bracket_App}
D(f, g)=P^{\mu\nu}\partial_\mu f \partial_\nu g +f X^\mu \partial_\mu g - g X^\mu \partial_\mu f \ ,
\ee
We require $D(f,g)$ to concide with our bracket
\be 
\left\{ f,g\right\} _{\zeta}=\left\{ f,g\right\} +c_\zeta \,\zeta(u+v) \left(f\partial_{w}g-g\partial_wf\right) \ ,
\ee 
with the first term defined in \eqref{poisson_def}. We can then read off
\be\label{P_and_X_explicit}
P^{\mu\nu}=\left\{\begin{matrix}
-1,&\mu=u,\ \nu=w \\[-5pt]
\ \ 1,&\mu=w,\ \nu=u \\[-5pt] 
\ \ 0,&\text{otherwise} 
\end{matrix}\right.  \qquad \text{and} \qquad
X^\mu=\left\{\begin{matrix}
c_\zeta \,\zeta(u+v), &\mu=w \\[-5pt] 
0,&\text{otherwise} 
\end{matrix}\right.
\ee
The Jacobi bracket is required to satisfy the conditions in \eqref{conds_Jacobi}:
\be \label{app:conds_jac}
[P, P]=2 X \wedge P \ , \quad [X, P]=0 \ .
\ee
They ensure that the Jacobi identity is satisfied. Thus we could check them indirectly, by looking at the Jacobi identity, but let us write them explicity, as a sanity check that we have correctly identified $P^{\mu\nu}$ and $X^\mu$. The Schouten-Nijenhuis bracket between an m-tensor $A$ and a p-tensor $B$ is given by:
\be 
[A,B]^{\mu_1...\mu_{m+p}}=\tfrac{1}{(m-1)!p!}\varepsilon^{\mu_1...\mu_{m+p}}_{\nu_2...\nu_m\rho_1...\rho_p}
A^{\sigma\nu_2...\nu_m}\frac{\partial B^{\rho_1...\rho_p}}{\partial x^\sigma}
+\tfrac{1}{m!(p-1)!}\varepsilon^{\mu_1...\mu_{m+p}}_{\nu_1...\nu_m\rho_2...\rho_p}B^{\sigma\rho_2...\rho_p}
\frac{\partial A^{\nu_1...\nu_m}}{\partial x^\sigma}
\ee 
Since all the components of $P^{\mu\nu}$ are constant, we immediately have
\be 
[P,P]=0
\ee 
We then explicitly have 
\be 
2 X\wedge P= 4 c_\zeta \,\zeta(u+v) \frac{\partial}{\partial w}\wedge \frac{\partial}{\partial w} \wedge \frac{\partial}{\partial u}
=0 \ ,
\ee 
thus the first condition is immediately satisfied. An alternative way to see the above is by noting that $P$ and $X$ can be seen as tensors in  a two-dimensional space spanned by $u$ and $w$ ($P$ is simply the epsilon tensor in this space), where we treat $v$ and $\bar{w}$ as parameters. Then the first equation in \eqref{app:conds_jac} is trivially satisfied, as both sides vanish, since they are three-forms. 

For the second equality in \eqref{app:conds_jac} we can write explicitly
\be 
[X,P]^{\mu_2\mu_3}=\tfrac{1}{2}\varepsilon^{\mu_2\mu_3}_{\rho_1\rho_2}X^\sigma\frac{\partial P^{\rho_1\rho_2}}{\partial x^\sigma}+\varepsilon^{\mu_2\mu_3}_{\nu\rho_2}P^{\sigma\rho_2}\frac{\partial X^\nu}{\partial x^\sigma}
=c_\zeta \, \varepsilon_{wu}^{\mu_2\mu_3}\frac{\partial\zeta(u+v)}{\partial w}=0
\ee 
where we additionally made use of the fact that the non-zero coefficient of $X$ is independent of $w$.

%%%%%%%%%%%%%%%%
\section{Properties of conformally self-dual metrics}
\label{ap:ConfSD}
We proceed to show explicitly the expressions for the energy density and equation of state of the power law cosmological self-dual solutions.
\paragraph{Conformally flat self-dual}
For conformally flat self-dual solutions, the energy density is
\begin{equation}
\rho=\frac{M_{Pl}^2}{L^2}a^{-\frac{2(1+p)}{p}}p\left(3p+\left(3p+\hat{\Theta}_\text{flat}\right)\partial^2_w\phi\right) \ ,
\end{equation}
where $a=(u+v)/(2 L)$ and the operator $\hat{\Theta}_\text{flat}$ is
\begin{equation}
\hat{\Theta}_\text{flat}=\frac{1+p}{2}\partial^2_w\phi-\tau(\partial_\tau+2(\partial_u\partial_w\phi)\partial_w) \ ,
\end{equation}
and the equation of state is given by
\begin{equation}
\omega=\frac{\tilde{P}}{\tilde{\rho}}=\frac{2-p}{3p}+\frac{2(p-1)}{3p}\frac{\hat{\Theta}_\text{flat}\partial^2_w\phi}{3p+\left(3p+\hat{\Theta}_\text{flat}\right)\partial^2_w\phi} \ .
\end{equation}
Thus, we obtain the usual FLRW equation of state when the second term in the equation above vanishes. For a generic $\phi$, this is only satisfied if $p=1$, which corresponds to the case of the self-dual radiation solution discussed in Section \ref{sec:radiation}. 

Alternatively, as long as the second term is small, the solution has an equation of state that approximates the corresponding perfect fluid. One possibility is to have $\partial^2_w\phi\ll1$, in which case the equation of motion of the scalar reduces to $\square_{\mathbb{R}^{4}}\phi-(\partial_u\partial_w\phi)^2\simeq0$. 

%%%%%%%%%%%%%%%%
\paragraph{Conformally dS self-dual}
In the case of metrics conformal to the dS self-dual solution, we use $\Omega=(\tau/L)^{p+1}$ in Eq.~\eqref{eq:Conf} and find that the energy density is given by
\begin{equation}
    \rho=\frac{M_{Pl}^2}{L^2}a^{-\frac{2(1+p)}{p}}\left(3p^2+(1+p)\left((1+3p)+\hat{\Theta}_\text{dS}\right)\partial^2_w\phi\right) \ , 
\end{equation}
where the operator $ \hat{\Theta}_\text{dS}$ is 
\begin{equation}
    \hat{\Theta}_\text{dS}=\frac{1}{2}p\partial^2_w\phi+2(\partial_w\phi)\partial_w-\tau(\partial_\tau+2(\partial_u\partial_w\phi)\partial_w) \ , 
\end{equation}
and the equation of state reads
\begin{equation}
 \omega=\frac{1}{3}\left(\frac{\frac{3p(2-p)}{(1+p)}+\left((1-3p)+\hat{\Theta}_\text{dS}\right)\partial^2_w\phi}{\frac{3p^2}{(1+p)}+\left((1+3p)+\hat{\Theta}_\text{dS}\right)\partial^2_w\phi}\right)   \ .
\end{equation}
From this, we can see that when $p=0$, we recover the equation of state of radiation. For the case $p\neq0$, we can rewrite the equation of state as
\begin{equation}
\omega=\frac{2-p}{3p}-\frac{2\left((1+2p)+(1-p)\hat{\Theta}_\text{dS}\right)\partial^2_w\phi}{3p\left(\frac{3p^2}{1+p}+\left((1+3p)+\hat{\Theta}_\text{dS})\right)\partial^2_w\phi\right)} \ .
\end{equation}
which shows that, just like in the conformally flat self-dual case, we can approach the FLRW equation of state if $\partial^2_w\phi\ll1$. 

%%%%%%%%%%%%%%%%%
\paragraph{Conformally coasting self-dual}
In this last case, we take $\Omega=e^{-\tau/L}\tau^{p}$ in Eq.~\eqref{eq:Conf} to obtain power law scale factors in conformal time. With this choice, we find that the energy density is given by
\begin{equation}
\rho=\frac{M_{Pl}^2}{L^2}a^{-\frac{2(1+p)}{p}}\left(3p^2+\left(3p^2+\hat{\Theta}_\text{non-a.}\right)\partial^2_w\phi\right)\ , 
\end{equation}
where the operator $ \hat{\Theta}_\text{non-a.}$ is 
\eqs{
\hat{\Theta}_\text{non-a.}=&p(1+p)\partial^2_w\phi /2\nonumber \\
&-p\tau(\mathcal{H}\left(2+\partial_w\phi\partial_w\right)+\partial_\tau+2(\partial_u\partial_w\phi)\partial_w) \nonumber \\
&\tau^2\left(-\mathcal{H}^2\partial^2_w\phi/2+\mathcal{H}\left(\partial_\tau+(\partial_u\partial_w\phi)\partial_w\right)\right)\ .
}
Meanwhile, the equation of state can be written as 
\begin{equation}
 \omega=\frac{3p(2-p)+\left(3p(2-p)+6 p \mathcal{H}\tau+\hat{\Theta}_\text{non-a.}\right)\partial^2_w\phi}{3\left(3p^2+\left(3p^2+\hat{\Theta}_\text{non-a.}\right)\partial^2_w\phi\right)}   \ ,
\end{equation}
which, as in the conformally dS self-dual case, reduces to $1/3$ for $p=0$. When $p\neq0$, we can write
\begin{equation}
\omega=\frac{2-p}{3p}+\frac{1}{3}\left(\frac{\left(6 p \mathcal{H} t+\frac{2(p-1)}{p}\hat{\Theta}_\text{non-a.}\right)\partial^2_w\phi}{3p^2+\left(3p^2+\hat{\Theta}_\text{non-a.}\right)\partial^2_w\phi}\right)  \ ,
\end{equation}
such that, like in all the previous cases, when $\partial^2_w\phi\ll1$, the equation of state approaches that of the usual FLRW spacetime.

%%%%%%%%%%%%%%%%
\bibliographystyle{JHEP}
\bibliography{references}
%%%%%%%%%%%%%%%%
\end{document}